# *Science Objectives for an X-Ray Microcalorimeter Observing the Sun*

J. Martin Laming[1]§, J. Adams[3], D. Alexander[8], M Aschwanden[12], C. Bailey[3], S. Bandler[3], J. Bookbinder[2], S. Bradshaw[8], N. Brickhouse[2], J. Chervenak[3], S. Christe[3], J. Cirtain[3], S. Cranmer[2], S. Deiker[12], E. DeLuca[2], G. Del Zanna[13], B. Dennis[3], G. Doschek[1], M. Eckart[3], A. Fludra[15], F. Finkbeiner[3], P. Grigis[2], R. Harrison[15], L. Ji[2], C. Kankelborg[14], V. Kashyap[2], D. Kelly[3], R. Kelley[3], C. Kilbourne[3], J. Klimchuk[3], Y.-K. Ko[1], E. Landi[6], M. Linton[1], D. Longcope[14], V. Lukin[1], J. Mariska[1], D. Martinez-Galarce[12], H. Mason[13], D. McKenzie[14], R. Osten[17], G. Peres[10], A. Pevtsov[16], K. Phillips[4] F. S. Porter[3], D. Rabin[3], C. Rakowski[1], J. Raymond[2], F. Reale[10], K. Reeves[2], J. Sadleir[3], D. Savin[11], J. Schmelz[9], R. K. Smith[2], S. Smith[3], R. Stern[12], J. Sylwester[5], D. Tripathi[13], I. Ugarte-Urra[7], P. Young[7], H. Warren[1], B. Wood[1]

*[1]Naval Research Laboratory, [2]Smithsonian Astrophysical Observatory, [3]NASA Goddard Space Flight Center, [4]Mullard Space Science Laboratory, UCL, [5]Space Research Centre, Polish Academy of Sciences, [6]University of Michigan, [7]George Mason University, [8]Rice University, [9]University of Memphis, [10]University of Palermo, [11]Columbia University, [12]Lockheed Martin, [13]DAMTP, University of Cambridge,[14]Montana State University, [15]STFC Rutherford Appleton Laboratory,[16]National Solar Observatory, [17]Space Telescope Science Institute*

**ABSTRACT**

We present the science case for a broadband X-ray imager with high-resolution spectroscopy, including simulations of X-ray spectral diagnostics of both active regions and solar flares. This is part of a trilogy of white papers discussing science, instrument (Bandler et al. 2010), and missions (Bookbinder et al. 2010) to exploit major advances recently made in transition-edge sensor (TES) detector technology that enable resolution better than 2 eV in an array that can handle high count rates. Combined with a modest X-ray mirror, this instrument would combine arcsecond-scale imaging with high-resolution spectra over a field of view sufficiently large for the study of active regions and flares, enabling a wide range of studies such as the detection of microheating in active regions, ion-resolved velocity flows, and the presence of non-thermal electrons in hot plasmas. It would also enable more direct comparisons between solar and stellar soft X-ray spectra, a waveband in which (unusually) we currently have much better stellar data than we do of the Sun.

## 1. THE PROMISE OF X-RAY MICROCALORIMETERS IN SOLAR PHYSICS

Coronal heating has been the central problem of the solar outer atmosphere for over half a century (e.g. Klimchuk 2006). The advent of high quality imaging data in the last 15 years (SOHO, TRACE, STEREO, Hinode, and now SDO) has forced the realization that the solar atmosphere is significantly more dynamic and turbulent than had been previously appreciated. Since much of this behavior occurs on few seconds to minutes timescales, comparable to the integration time to record a spectrum or to scan a raster to make an image with grating instruments, what is ideally needed is an instrument capable of imaging spectroscopy, with high throughput, and high spatial and spectral resolution. In this respect, recent advances in microcalorimeter arrays for solar X-ray observations have the potential to revolutionize our understanding of coronal plasmas, specifically with regard to the non-equilibrium plasma dynamics of solar flares and active regions. The microcalorimeter is well-matched to the parameters of interest for detailed studies of the fundamental plasma physics in the Sun's atmosphere, with excellent spectral (<2 eV), spatial (~arcsec) and temporal (millisecond) scales and broad spectral coverage from 0.2 to 10 keV. Such an instrument will be able to:

1. Detect and resolve hundreds of atomic transition lines from different ionization states of many elements in the solar atmosphere

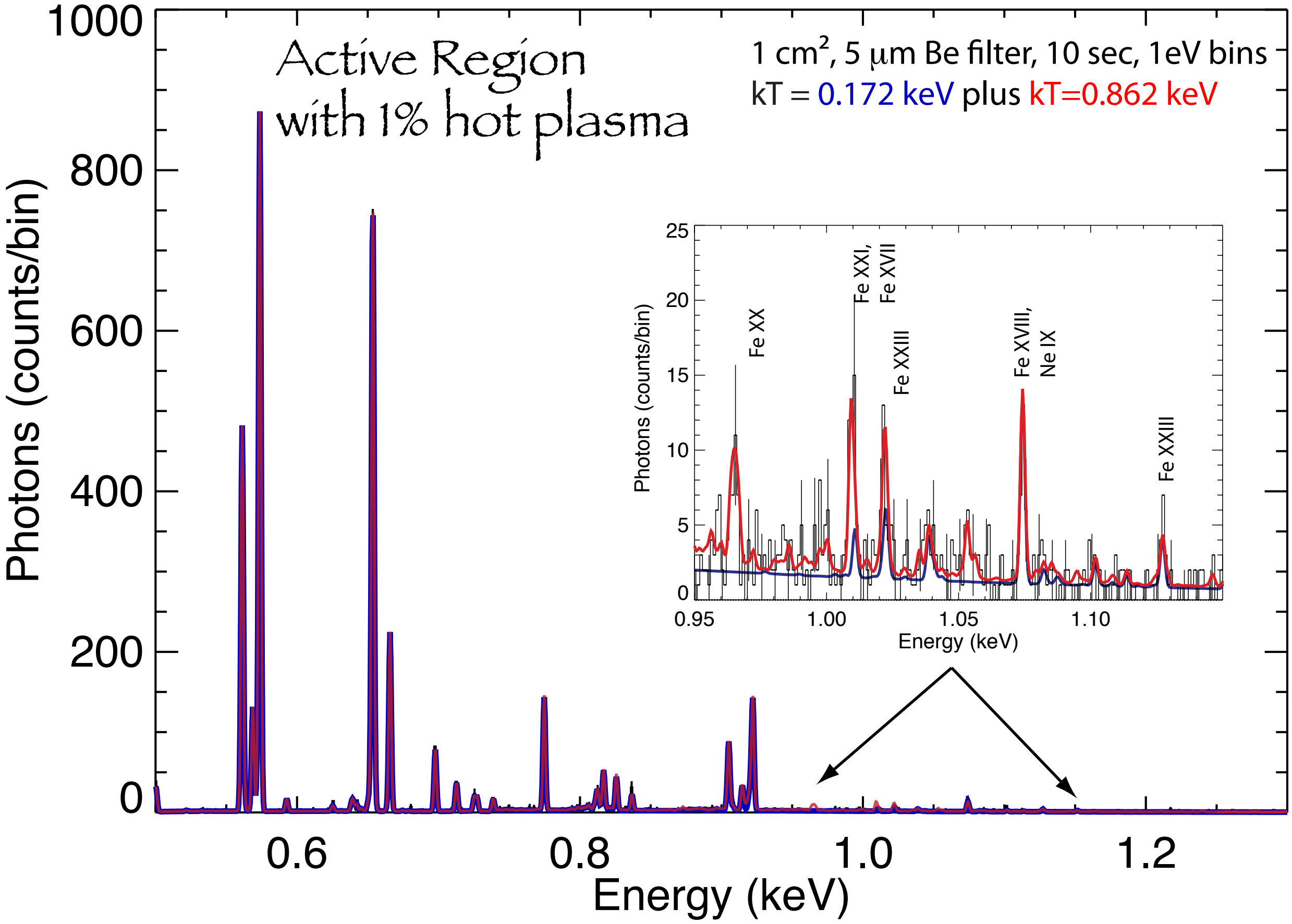

*Figure 1 – The spectrum shown here is from a plasma at 2 MK combined with a 10 MK component whose emission measure is 1% of the dominant term (black is the simulated combined data; blue the cool phase only, red the combined model). The zoomed-in region inset shows strong lines of higher charge state iron ions that are characteristic of the higher temperature region. Note that this spectrum is from one square arcsecond, so spatial variations in the hotter plasma could easily be imaged with this technique.*

2. Determine thermal and non-thermal electron distributions from lines and continuum
3. Determine plasma densities from ~$10^9$–$10^{13}$ cm$^{-3}$ and temperatures from <0.7 to >10 MK
4. Detect flows with velocities down to < 50 km s$^{-1}$
5. Determine the abundances of elements from carbon to nickel relative to hydrogen

The simultaneous combination of high resolution spectroscopic imaging observations across this entire energy range is new and tremendously powerful. Solar microcalorimeters will allow all of this information to be obtained simultaneously on time scales that allow us to follow the evolution of the radiative emission during the impulsive phase of flares. This is in contrast to the several minute cadence of previous instruments that sample only a narrow part of the spectrum and require rastering to build up images. The unique capabilities of these detectors will allow us to study how magnetic energy is released in magnetic reconnection events such as nano-flare heating in active regions, solar flares and, with different optics, coronal mass ejections with unprecedented spectral resolution.

With a 1-2 cm$^2$ mirror and a thin Be filter, a typical active region or C flare region can be observed at full resolution. Gradual spectral resolution degradation will occur at increased incident flux. Nonetheless, brighter flares, up to and including X class flares, can be observed at full spec-tral resolution using a combination of thicker filters (which remove the less-useful soft photons) and possibly a neutral density filter.

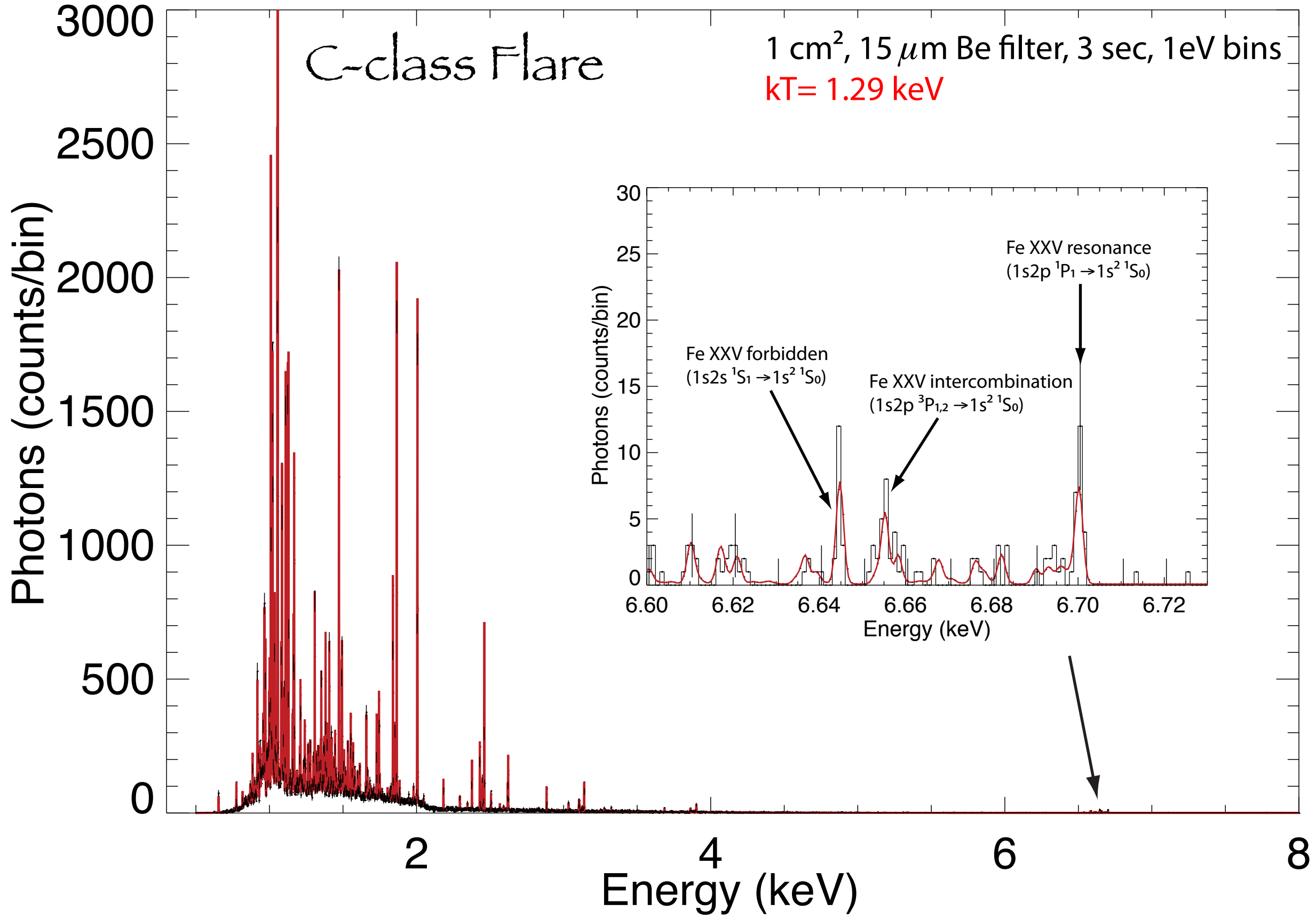

*Figure 2 – The full spectrum shows the large number of lines available in a single 3 second observation of one square arcsecond, primarily from the iron L-shell ions combined with the H-and He-like lines of abundant ions. The inset is a blown up segment showing the resolved Fe XXV line complex with the red curve showing the model and the black the simulated data.*

## 2. SOLAR ACTIVE REGIONS

The broadband response of a microcalorimeter is ideally suited to explore modern theories of coronal heating. Following Parker's (1988) initial suggestion, the idea that the corona is heated by short intense releases of energy, commonly known as "nanoflares" or "microflares" has gained currency. A number of indirect observations support this scenario. The scale heights of coronal loops are clearly incompatible with static models. Recent models of the FIP effect[1] where the ponderomotive force from Alfvén waves drives the fractionation (Laming 2004b; 2009) also strongly suggest a coronal source for these waves, again more compatible with impulsive rather than quasi-static coronal heating. Amidst this, the classical "smoking gun", the direct detection of hot plasma (with a low emission measure) signalling the location(s) of the heating event, has remained elusive.

Tantalizing hints are seen when the solar corona is imaged with Bragg Crystal Spectrom-eters in the light of Mg XII 8.42 Å (Zhitnik et al. 2006, Urnov et al. 2007), in emission measure analyses of spectra taken with Hinode/EIS (Patsourakos & Klimchuk 2009)[2], with Hinode/XRT using different filter ratios (Schmelz et al. 2009, Reale et al. 2009a) and/or in quiet sun observations with RHESSI (McTiernan 2009, Reale et al. 2009b). On the other hand Warren et al. (2010) study the Hinode/EIS spectrum from the core of a solar active region and find no evidence for very hot plasma which presents a challenge to the theory of nanoflares. The observational evidence to date all suffers from the use of either filter ratios with a broad band telescope, where low temperature emission has to be corrected for, or the observation of an intrinsically weak signal at low spatial resolution. The imaging capabilities of a microcalorimeter array, combined with its broad bandpass and dynamic range, will allow detection of small admixtures of recently-heated plasma in a region otherwise dominated by cooler material.

*[1]The coronal enhancement in abundance of elements with First Ionization Potential (FIP) less than about 10 eV, i.e. Mg, Si, Fe, over their photospheric values, while elements with higher FIP remain essentially unchanged.*

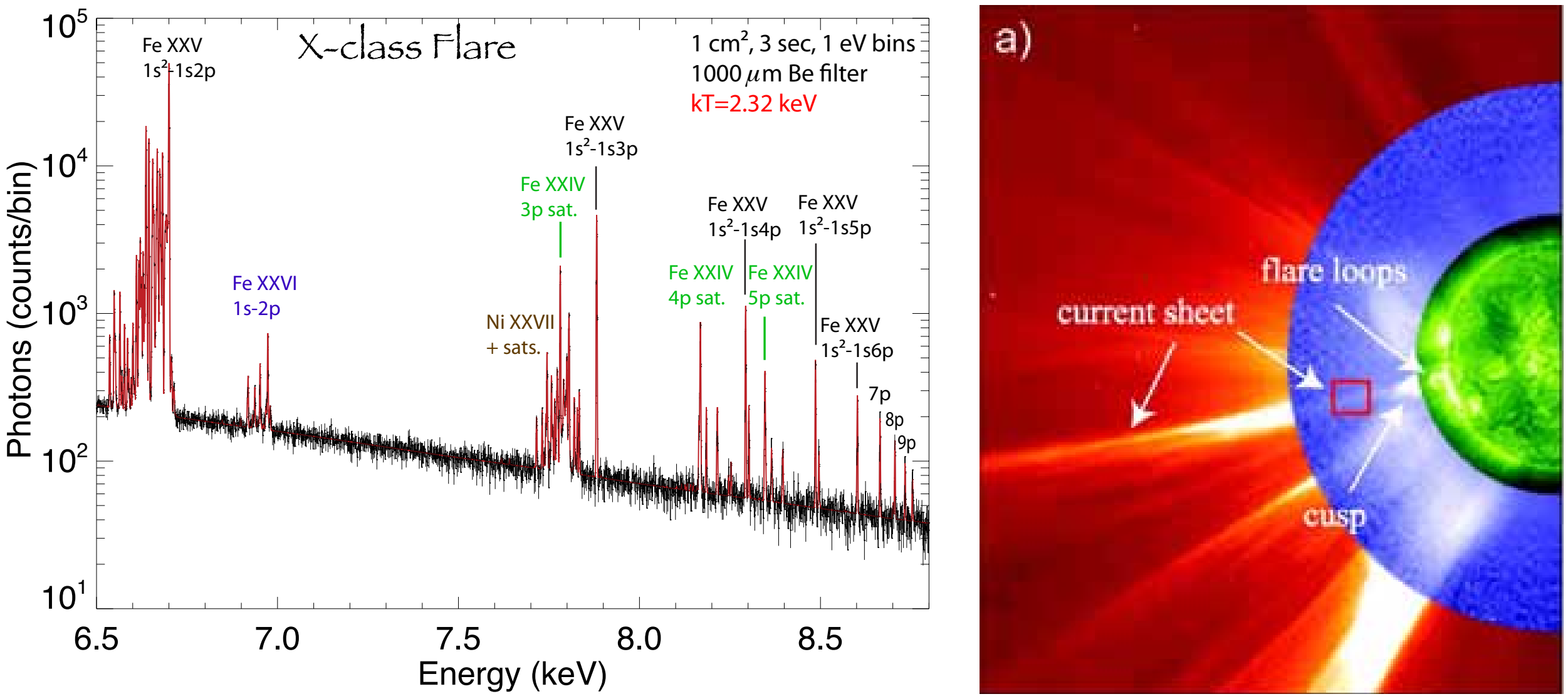

*Figure 3 – Left: Theoretical solar flare spectrum, showing the Fe (6.7 keV) and Fe/Ni (8 keV) spectrum at a few eV resolution. RHESSI observes these Fe and Ni lines as single broad features at 1 keV resolution. The continuum is free-free and free-bound. The assumed temperature is 20MK (typical of medium-to-large flare peak temperatures) Right: Composite LASCO/C2, Mauna Loa/Mark IV, and EIT/Fe XII image showing the current sheet related to the CME that occurred on 2003 November 18 (taken from Bemporad 2008).*

Fig. 1 shows two views of a simulated single pixel calorimeter spectrum of a solar active region. The imaged emission measure is $4\times10^{38}$ cm$^{-3}$ at $2\times10^{6}$K, with a small contribution $4\times10^{36}$ cm$^{-3}$ at $10^{7}$ K, to model the anticipated nanoflare component. The left panel shows the broad band 0.5-1.3 keV region. The right panel shows a blow-up of the 0.95-1.15 keV region, and the number of hot lines (Fe XVII, XVIII, XX, XXI, XXIII) that immediately become visible. Outside this region the Mg XII 8.42Å Lyman α transition already used (Zhitnik et al. 2006, Urnov et al. 2007) becomes available, along with other He- and H-like systems.

The microcalorimeter resolution < 2 eV means that Lyman α from H-like ions Mg XII and up will be resolved, and the intensity ratio between the two components will be measurable. Several of these ratios varying from 2:1 have been previously detected by instruments on solar flare missions (P78-1, Solar Maximum Mission, Hinotori, all without any imaging or other context instrumentation), and generally interpreted as due to effects of radiation transfer in an asymmetrical plasma, such that lines with higher opacity escape preferentially in one direction; we note this is easily tested with a calorimeter via observations of higher Lyman series lines. Laming (1990) suggested that an alternative explanation could lie with directed electrons, presumably accelerated along the magnetic field direction, which impart some polarization to the levels they excite. Upon radiative decay, the polarized light is naturally emitted in an anisotropic fashion, giving rise to different Lyman α intensity ratio for different angles of observation with respect to the "beam" axis. The relaxation time of ~keV electrons in coronal plasma is sufficiently short to make this unlikely in the case of a flare produced electron beam; the suprathermal electrons would have to be continuously replenished, however a nanoflare model of coronal heating might be expected to yield just such a signature, especially if the heating events are associated with small scale magnetic reconnection.

[2]*Lines from hot plasma also appear in the EUV, but are weak and difficult to isolate from other emission from lower temperatures.*

Accompanying the energy release in the corona by nanoflares should be episodes of evaporation from the chromosphere, following heat conduction downwards. This upflow is predicted to be 50-100 km $s^{-1}$, i.e. detectable with the microcalorimeter. To date, such upflows have generally only been convincingly detected during the impulsive phase of solar flares (e.g. Milligan et al. 2006ab) as a "shoulder" on the blue wing of an otherwise stationary line profile. Czaykowska et al. (1999) see an upflow of 50 km $s^{-1}$ during the late gradual flare phase, through the shift of the entire line profile.

## 3. SOLAR FLARES

Solar flares have traditionally been observed with Bragg Crystal Spectrometers flown on missions at solar maximum around 1980 (P78-1, Solar Maximum Mission, Hinotori) and again around 1991 (Yohkoh). The microcalorimeter brings all of the attributes of these missions together with broad bandpass and X-ray imaging. The higher throughput compared to earlier missions will allow observations of the early and preflare stage of the event, before the hard X-ray burst. Battaglia et al. (2009) argue that this early emission stems from conduction-driven chromospheric evaporation following heating in the corona. With a spectral resolution of < 2 eV, the thermal Doppler widths of Fe lines in evaporating plasma would be easily resolved; most likely there is always also a "non-thermal" or turbulent component to the line width but it appears to be < 2 eV, from SMM BCS spectra of the main phase of flares (it is much larger at the flare impulsive stage, e.g. Alexander et al. 1998).

During this phase it would also be key to look for non-ionization equilibrium effects through line ratios in particularly the 6.7 keV Fe line complex which is made up of He-like Fe (Fe XXV) lines (resonance, intercombination, forbidden) and Fe XXIV and lower ionization stage dielectronic satellites. Intensity ratios of satellites formed by dielectronic recombination and the resonance line give electron temperature, those formed by inner-shell excitation give the Li-like/He-like Fe ion abundance ratio. If there is ionization equilibrium, the temperatures from the two ratios should be equal but non-equality would indicate either an ionizing or recombining plasma. Fig. 2 shows a simulated spectrum from one calorimeter pixel of a C class flare, obtained in 3 sec integration.

The 6.7 keV "Fe line" complex consists of many closely spaced lines. However, the higher-excitation 8 keV "Fe/Ni line" feature (see Phillips 2004), consisting mostly of Fe XXV $1s^2 - 1snp$ lines (n = 3,4,5...) and satellites which crowd together to form single spectral features, is easier to resolve, and would therefore be sensitive to lower upflow velocities. This feature has been observed with solid-state detectors (e.g. on the NEAR spacecraft, resolution ~600 eV) and with RHESSI (1 keV resolution) but never with crystal spectrometers, in space or in the laboratory. The left panel of Fig. 3 shows a simulation of this spectral region. Ratios of satellites/parent (Fe XXV) lines give electron temperatures and possibly departures from thermal distributions. Evidence exists for high (approaching $10^{12}$–$10^{13}$ cm$^{-3}$) electron densities in solar flares (Phillips et al. 1996), using diagnostic line ratios in Fe XXI and Fe XXII observed with crystal spectrometers on SMM. The microcalorimeter would provide similar spectral resolution, but with significantly higher bandpass, throughput and spatial resolution. Hence the spatial variation of several different density diagnostic line ratios could be evaluated, with potentially important consequences for our understanding of the physics of solar flares.

## 4. OFF-LIMB OBSERVATIONS OF CMES AND THE SOLAR WIND

Off-limb solar observations will be perhaps where the combination of microcalorimeter attributes finds its optimum application. Observations of disk flares have revealed large blue shifts at onset in $10^7$ K plasma (Innes et al. 2001, Innes et al. 2003), reaching as high as 650 - 1000 km s$^{-1}$. These velocities are similar to the Alfvén speed expected, and so could be the Alfvénic outflow from the reconnection region, or due to turbulence. Off limb current sheets have been imaged with Hinode/XRT (Savage et al. 2010), and modeled (Reeves et al. 2010). Abundances have been previously measured in CME current sheets (Ciaravella et al. 2002, Ko et al. 2003), with the result that the FIP effect is present and strong. Microcalorimeter observations should make substantial progress on outstanding unsolved issues surrounding magnetic reconnection; the partitioning of energy between thermal energies of electrons and ions, bulk kinetic energy and turbulence.

Evidence is also mounting that CMEs must be heated as they erupt, and that this heating represents a significant fraction of the CME energy budget. Analyses of ion charge states detected in situ (Rakowski et al. 2007), and of remotely sensed UV emission (Akmal et al. 2001; Ciaravella et al. 2001; Lee et al. 2009; Landi et al. 2010) reveal that previously cold (i.e. ~$10^6$ K coronal material or cooler) is heated to temperatures approaching $10^7$ K as the filament erupts, and microcalorimeter observations ought to be able to detect such hot emission.

The shock wave driven ahead of the CME would also make an interesting target. Spectroscopic diagnostics exist within the calorimeter bandpass for the measurement of postshock electron temperature (the "G" ratio of He-like C V, N VI, O VII, etc). Ion temperatures can be measured from the line broadening, especially for shocks approaching 1000 km s$^{-1}$. The derived electron and ion internal energies should add up to the total shock energy (derived from kinematics), unless significant energy is being lost to the acceleration of solar energetic particles.

There is also evidence that the fast solar wind is heated in the region between about 1.5 and 5 $R_\odot$ heliocentric distance where it is accelerated, from analyses of ion charge states collected *in situ* (Laming 2004a, Laming & Lepri 2007) and the observations of optical and near IR lines from different Fe charge states during eclipses (Habbal et al. 2007; 2010). A microcalorimeter should be able to detect C V, VI, N VI, VII, and O VII, VIII in this region as confirmation of these inferences, and to compare with *in situ* observations. Although the electron temperature in this region may be as high as a few $\times 10^6$ K, Fe charges remain those of M-shell ions, because ionization equilibrium is rapidly lost in the low-density solar wind plasma.

## 5. CONCLUSIONS

While great progress has been made from X-ray imaging and UV and EUV imaging/spectroscopy, in the next decade we should return to the natural spectral band — the X-ray band — for studying high energy coronal phenomena in detail. This offers potentially decisive contributions to the emerging nanoflare model of coronal heating, through the detection of hot plasma associated with the heating. The microcalorimeter is also ideally suited to observations off-limb, and appears poised to advance this new field as well. The combination of spectral, spatial and temporal resolution over a broad spectral band afforded *simultaneously* by the microcalorimeter offers a major advance in diagnostic capability. Current instruments generally give either spatial or temporal resolution, and then over a relatively narrow spectral bandpass if high spectral resolution is required. The broad bandpass will also allow the acquisition of solar spectra that are much more comparable to stellar X-ray spectra returned by instruments on satellites like Chandra and XMM-Newton. Currently, we are in the unusual situation of having better stellar spectra in this spectral region than we do of the Sun (Osten 2010); a microcalorimeter will redress this mismatch, enabling direct comparisons with stellar observations in addition to addressing the issues in solar physics discussed above.